\documentclass[%
 aip,
cp,  % Conference Proceedings
 amsmath,amssymb,%nobibnotes,
 reprint,%
]{revtex4-2}

\usepackage{graphicx}% Include figure files
\usepackage{dcolumn}% Align table columns on decimal point
\usepackage{bm}% bold math
\usepackage[utf8]{inputenc}
\usepackage[T1]{fontenc}
\usepackage{mathptmx}
\begin{document}

\title{New methods for analytical calculation of elliptic \\
integrals, applied in various physical problems}
\author{Bogdan G. Dimitrov} % Write as First name Surname
 \email[Corresponding author: ]{dimitrov.bogdan.bogdan@gmail.com}
%\author{Author's Name}%
 %\email{second.author@institution.edu.}
\affiliation{
  Institute of Nuclear Research and Nuclear Energetics,
Bulgarian Academy of Sciences, 72 Tzarigradsko shaussee, 1784 Sofia,
Bulgaria.% Force line breaks with \\ if necessary
}

%\author{Another's Name}
 %\email{third.author@anotherinstitution.edu}
\affiliation{%
Institute for Advanced Physical Studies , Sofia Tech Park, 111
Tzarigradsko shaussee, 1784 Sofia, Bulgaria  \\e-mail:
bogdan.dimitrov@iaps.institute}% Force line breaks with \\ if necessary

%\affiliation{You would list an author's second affiliation (if applicable) here.}

\date{\today} % It is always \today, today, but any date may be explicitly specified
              % Not printed for conference proceedings

\begin{abstract}
{\itshape} A short review will be made of elliptic integrals, widely applied
in GPS (Global Positioning System) communications (accounting for General Relativity Theory-effects), cosmology, Black hole
physics and celestial mechanics. Then a novel
analytical method for calculation of zero-order elliptic integrals in the
Legendre form will be presented, based on the combination of several methods
from the theory of elliptic functions: 1. the recurrent system of equations
for higher-order elliptic integrals in two different representations. 2.
uniformization of four-dimensional algebraic equations by means of the
Weierstrass elliptic function 3.a variable transformation, inversely
(quadratically) proportional to a new variable. The developed method is a step 
forward towards constructing analytical methods, which 
can improve the precision of the calculation of elliptic integrals, necessary both for 
theoretical and experimental problems.  
\end{abstract}

\maketitle

\section{\label{sec:Intro}Introduction}

Elliptic integrals appear in various problems in cosmology, General
Relativity Theory, Black Hole physics, theoretical mechanics (Eulers
equations for rotation of a rigid body), integrable systems and etc. In the
current literature, two major facts remain unexplained: 1. Why various types
of elliptic integrals appear so frequently? In this paper, the answer to this fundamental problem 
will not be given. Moreover, the mentioned elliptic integrals in this paper are only a small part of the many other 
elliptic integrals, appearing in other seemingly distant areas of physics - theory of the rotation of the Earth
(accounting for the approximation of the Earth with a homogeneous ellipsoid), changes of the orbital period 
and eccentricities in binary pulsars, caused by the emission of gravitational waves and etc. An attempt to 
systematize as many as possible elliptic integrals will be made in another publication. It can only be supposed that 
one of the reasons for the frequent appearance of elliptic integrals is hidden in the complicated, nonlinear 
structure of the resulting equations. 2. The second important problem is how the approximations used
in the evaluation of elliptic integrals are related to the accuracy of the
obtained numerical results? \ Here one should mention that many authors make use of  
numerical methods for calculation of elliptic integrals, which is natural from the 
point of view about the necessity for obtaining concrete numbers. 
 It is also important to mention that in his book \cite{Legendre:1825} Legendre has not proved that
elliptic integrals of the first kind cannot be calculated analytically. Further in the 
text, some examples will be given, concerning authors, who imply analytical methods for calculation of 
elliptic integrals.  

In a previous publication \cite{DimitrovAIP:2022}, it was shown that when calculating the
propagation time of a signal, emitted by a GPS-satellite on a plane
elliptical orbit, the propagation time can be  expressed by a
combination of elliptic integrals of the first, second and the third rank.
Moreover, for the case of space-distributed orbits, second-order and
third-order elliptic integrals can be given also by an analytical
formulae.

The last fact naturally raises the question about the implementation of specific 
analytical algorithms for calculation of elliptic integrals. By "specific" it is meant that 
these methods are not the typical ones for the numerical  calculation of indefinite and 
definite integrals (see for example Ch. 4 of the monograph \cite{Scherer:2017} about the 
brief summary of such methods) and more importantly, they are not directly related also to the Jacobi elliptic 
functions, widely used for expressing the solutions of various nonlinear evolution equations 
(the monograph \cite{Akhmediev:1997} and many other books). For the case of elliptic integrals in the Legendre 
form, it will be shown in this paper that the final expression for the elliptic integral might represent a 
rather complicated functions of the modulus parameter of the integral.   

 The aims of the present paper are the following: 1.
To provide concrete examples of elliptic integrals in various research areas
of physics, and to demonstrate that the rapid development of experimental
techniques requires the exact analytical solution of these integrals. In
fact, some authors already provide analytical solutions of elliptic
integrals. For example, in \cite{Kraniotis:2005} (some other papers by the
same author can also be seen) the resulting elliptic integrals for photonic
geodesics in a Kerr -(anti) de Sitter space-time are expressed in terms of
the Appells hypergeometric function.  \ 2. To present a new analytical
algorithm for solving the zero-order elliptic integrals in the Legendre
form, which will be based on the s.c. "uniformization of fourth-order algebraic
curves (of genus $1$)", developed in the monograph by Whittaker and Watson
\cite{WhittakerWatson:1927}.

The analytical treatment of the most commonly encountered elliptic integral
in the Legendre form raises the question whether the more general elliptic
integrals with a third- and a fourth-order polynomials under the square root
in the denominator can also be solved analytically. This will be briefly
commented in the next section.

\section{\label{sec:TypesElliptic}Types of elliptic
integrals in various physical \ problems}

     \  Since further in the next sections the proposed new methods will be purely mathematical,
in order to illustrate the possible applicability of the new approach and
also provide the motivation for the analytical approach from a physical
point of view, several examples will be given about some elliptic integrals.

\subsection{\label{sec:RelatPerihAdvance}Relativistic treatment of
perihelion advance in General Relativity}

\bigskip The first example concerns the relativistic advance of perihelion advance in
GR and the angle of deflection $\varphi $ of an orbit \cite{Weinberg:1972},
given by the integral
\begin{equation}
\varphi =\pm \int \frac{A^{\frac{1}{2}}(r)}{r^{2}\left( \frac{1}{J^{2}B(r)}-%
\frac{E}{J^{2}}-\frac{1}{r^{2}}\right) ^{\frac{1}{2}}}dr\text{ \ \ ,}
\label{A1}
\end{equation}%
where $A(r)$ and $B(r)$ are the expressions from the Einstein's static
isotropic metric
\begin{equation}
ds^{2}=B(r)dt^{2}-A(r)dr^{2}-r^{2}d\theta ^{2}-r^{2}\sin ^{2}\theta d\varphi
^{2}  \label{A2}
\end{equation}%
\begin{equation}
A(r)\equiv \left[ 1-\frac{2MG}{r}\right] \text{ \ ;\ \ }B(r)\equiv \left[ 1-%
\frac{2MG}{r}\right] ^{-1}\text{ \ \ .}  \label{A3}
\end{equation}%
The exact expression for Eq. (\ref{A1}) is
\begin{equation}
\varphi =\pm \int \frac{J}{r\sqrt{F(r)}}\pm 2MG\int \frac{dr}{r^{2}\sqrt{F(r)%
}}\text{ \ \ ,}  \label{A4}
\end{equation}%
where $F(r)$ is the quartic polynomial
\begin{equation}
F(r)\equiv -EJr^{4}+2MGEJr^{3}+(1-J^{2})r^{2}+\left( 2MGJ^{2}-4MG\right)
r+(2MG)^{2}\text{ \ \ .\ }  \label{A5}
\end{equation}%
In accord with the definitions, which shall be given further, Eq. (\ref{A4})
represents a sum of two elliptic integrals of the third kind \ and of the
first and second order respectively. The formulae is solved analytically in
the first approximation of $\frac{GM}{r_{\pm }}$ as $\Delta \varphi =6\pi
\frac{MG}{L}=43.03^{^{\prime \prime }}$ ($\Delta \varphi =43.11\pm
0.45^{^{\prime \prime }}$for a century, according to the monograph \cite%
{Weinberg:1972}, written in 1972).Yet, there is no explanation why there is
a slight difference between the numbers.

\subsection{\label{sec:LightDeflSchwarzsch}Light deflection in the
Schwarzschild geometry (the case of non-rotating black holes)}

This second example about the angle difference between a falling and a
deflected light ray is closely related with the previous example. The angle
of deflection is given by \cite{GyulchevYazad:2017}
\begin{equation}
\phi _{m}-\phi _{\infty }=\int\limits_{0}^{1}\frac{dx}{\sqrt{1-x^{2}-\frac{%
2GM}{c^{2}r_{m}}(1-x^{3})}}\text{ \ \ ,}  \label{A6}
\end{equation}%

\bigskip where $\phi _{\infty }$ is the asymptotic value of $\phi $,
corresponding to the position of the photon at the time of its emission from
the light source. If distances $r_{m}$ for propagation of light rays are
considered, much greater than the gravitational radius of the lense $%
r_{m}\gg r_{s}=\frac{2GM}{c^{2}}$ (i.e. $h=\frac{r_{s}}{r_{m}}=\frac{2GM}{%
c^{2}r_{m}}$ is a small parameter), then the above integral can be computed
after decomposition of the denominator into an infinite sum with respect to
the small parameter $h$ and subsequent integrations. From an experimental
point of view, there is a necessity to be able to compute such integrals not
at infinity since VLBI observations with the submillimeter square radio
array (SMA) in Hawai, also the submillimeter radio observatory (CARMA) in
California have observed physical structures in the centre of Sgr A* with an
accuracy only $4$ $r_{s}$ and $5.5$ $r_{s}$ from the centre of the galaxy $%
M87$ \cite{GyulchevYazad:2017}. So these are distances not at infinity, consequently
exact analytical methods are needed and not approximate.

\subsection{\label{sec:LightKerrBH}Light trajectories around Kerr black holes%
}

The geodesic lines for the photon trajectories are represented by the
differential equations
\begin{equation}
\rho ^{2}\frac{dr}{d\lambda }=\pm \sqrt{R(r)}\text{ \ \ \ \ , \ \ \ \ }\rho
^{2}\frac{d\Theta }{d\lambda }=\pm \sqrt{\Theta (\theta )}  \label{A8}
\end{equation}%
where $R(r)$ is the polynomial
\begin{equation}
R(r)\equiv \left[ (r^{2}+a^{2})E-aL_{z}\right] ^{2}-\Delta \left[ \left(
L_{z}-aE\right) ^{2}+K\right] \text{ \ \ \ ,}  \label{A10}
\end{equation}%
$\Theta (\theta )$ is an expression, which here shall not be defined, $K$ is
the Carters constant. The left-hand side of the resulting from Eq. (\ref{A8}%
) expression after the integration \
\begin{equation}
\int \frac{dr}{\sqrt{R(r)}}=\pm \int \frac{d\Theta }{\sqrt{\Theta (\theta )%
}}  \label{A13}
\end{equation}%
represents an elliptic integral with a fourth-degree polynomial in the
denominator (fully consistent with the general mathematical definition of
elliptic integrals, which further shall be given, see also \cite%
{DimitrovAIP:2022}). The integral Eq. (\ref{A13}) is important for
calculating frame dragging for orbits of stars and periapsis advance for
orbits close to the event horizon of galactic black holes \cite%
{Kraniotis:2007}, also the deflection of light rays in the gravitational
field of a rotating mass (the Kerr field) \cite{Kraniotis:2005}.

\subsection{\label{sec:EllipCosmology}Elliptic integrals in cosmology,
apparent luminosity distance of supernovas and the accelerating Universe}

In cosmology, an important characteristic is the apparent luminosity
distance of a source $d_{L}(z)$ , observed at a redshift $z$. The apparent luminosity distance is
defined as \cite{Weinberg:2008}
\begin{equation}
d_{L}\equiv a(t_{0})r_{1}(1+z)=\frac{(1+z)}{H_{0}\Omega _{K}^{\frac{1}{2}}}%
\sinh \left[ \Omega _{K}^{\frac{1}{2}}M\right] \text{ \ \ \ \ ,}  \label{B4}
\end{equation}%
where $M$ is the zero-order elliptic integral with a fourth-degree
polynomial in the denominator
\begin{equation}
M\equiv \int\limits_{\frac{1}{1+z}}^{1}\frac{dx}{x^{2}\sqrt{\Omega
_{\Lambda }+\Omega _{K}\frac{1}{x^{2}}+\Omega _{M}\frac{1}{x^{3}}+\Omega _{R}%
\frac{1}{x^{4}}}}\text{ \ \ \ .}  \label{B5}
\end{equation}%
and $x$ is determined as $x\equiv \frac{a}{a_{0}}=\frac{1}{1+z}$. The total
density
\begin{equation}
\rho =\frac{3H_{0}^{2}}{8\pi G}\left[ \Omega _{\Lambda }+\Omega _{K}\left(
\frac{a_{0}}{a}\right) ^{2}+\Omega _{M}\left( \frac{a_{0}}{a}\right)
^{3}+\Omega _{R}\left( \frac{a_{0}}{a}\right) ^{4}\right]  \label{B2}
\end{equation}%
is the sum of the energy density $\frac{3H_{0}^{2}}{8\pi G}\Omega _{\Lambda
} $ in vacuum, the energy density of non-relativistic matter $\frac{3H_{0}^{2}}{8\pi G}\Omega
_{M} $, the energy density of relativistic matter (radiation) $\frac{3H_{0}^{2}}{8\pi G}\Omega
_{R}$ and of energy density $\frac{3H_{0}^{2}}{8\pi G}\Omega _{K}$ due to
the curvature of space-time, which is usually neglected if the Universe is
spatially flat $K=0$ (the equality $\Omega _{\Lambda }+\Omega _{K}+\Omega
_{M}+\Omega _{R}=1$ is also fulfilled, since $\Omega _{\Lambda }$, $\Omega
_{K}$, $\Omega _{M}$, $\Omega _{R}$ are normalized with respect to the \
critical energy density $\rho _{0,crit}=$ $\frac{3H_{0}^{2}}{8\pi G}%
=1.878.10^{-29}h^{2}$ $[g/sm^{3}]$). It is known that the
High-$z$ Supernova Search Team discovered $23$ new high redshift supernovas 
of type $Ia$, including $15$ supernova with $z>0.7$ and thus the best
fit-values for $\Omega _{\Lambda }$, $\Omega _{M}$ were found to be $\Omega
_{\Lambda }=0.67$ and $\Omega _{M}=0.33$ (meaning that $\Omega _{\Lambda
}>\Omega _{M}$ and $q_{0}=-\frac{1}{2}$, where $q_{0}$ is the deceleration
parameter, the minus sign indicating that the Universe is in a state of
accelerated expansion). Interestingly, for a flat Universe with $\Omega
_{K}=\Omega _{R}=0$ (which in fact follows from the normalization condition
and the values for $\ \Omega _{\Lambda }$ and $\Omega _{M}$ ($0.67+0.33=1.00$%
) , as noted in \cite{Weinberg:2008}, the existence of a galaxy with
redshift at $z=1.55$ $\ $and with age $\simeq 35$ $Gyr$ sets a lower bound
on the vacuum energy density $\Omega _{\Lambda }$ of about $0.6$. However,
the observationally determined value $\Omega _{\Lambda }=0.67$ raises the
question: is it necessary that the "curvature energy density" should also be
accounted? An estimate
\begin{equation}
\frac{a_{0}}{a}=\frac{\mid \Omega _{curv}\mid }{\Omega _{M}}>10\text{ \ \ \
, \ \ \ }\mid \Omega _{curv}\mid <0.02\text{ \ \ \ \ }  \label{B6}
\end{equation}%
has been performed in the monograph \cite{Rubakov:2017}, neglecting $\Omega
_{R}$ and $\Omega _{\Lambda }$ in the Friedmann equation (which however is
in contradiction with the observational data about the dominance of the
vacuum energy!). So the estimates should be performed on the base of the
Friedmann equation with all the terms included in formulae Eq. (\ref{B2}).

Therefore, the calculation of the elliptic integral Eq. (\ref{B5}) will help
to confirm that even in the presence of curvature, the apparent luminosity
distance of the type $Ia$ supernovae will also be greater than in the empty
model with $q_{0}=0$, meaning also that the apparent luminosity decreases
(i.e. supernovas will look fainter). In other words, since vacuum energy
has increased ($\Omega _{\Lambda }=0.67$), the luminosity distance has also
increased.

In this paper, elliptic integrals of the type Eq. (\ref{B5}), Eq. (\ref{A13}%
) and Eq. (\ref{A4}) with a polynomial of the fourth degree in the
denominator will not be considered. However, it will be suggested that the
same analytical algorithm, which concerns elliptic integrals with a
third-order polynomials in the denominator (in the Legendre form), can be
also applied with respect to these more general integrals.

\subsection{\label{sec:EllipticNonlinearEq}Elliptic integrals related to some
nonlinear equations}

The Korteweg-de-Fries equation is one of the examples about solutions in
terms of elliptic integrals, which have a relation with the approach in this
paper. If one sets up $u(x,t)$ $=u(\zeta =x-vt)$, then the simultaneous
account of dispersion and nonlinearity leads to the equation \cite%
{Martinson:2002}
\begin{equation}
-v\frac{du}{d\zeta }+u\frac{du}{d\zeta }+\beta \frac{d^{3}u}{d\zeta ^{3}}=0%
\text{ \ \ \ ,}  \label{B9}
\end{equation}%
which can be rewritten in the form
\begin{equation}
3\beta \left( \frac{du}{d\zeta }\right) ^{2}=F(u)\text{ \ \ \ ,}  \label{B10}
\end{equation}%
where $F(u)$ is the cubic polynomial
\begin{equation}
F(u)\equiv -u^{3}+3vu^{2}+6au+6b\text{ \ \ \ \ \ \ \ \ }a,b=const\text{ \ \ .%
}  \label{B11}
\end{equation}%
If all the three roots of the polynomial $F(u)\equiv
(u_{1}-u)(u_{2}-u)(u_{3}-u)$ are different, then the solution is given by
\cite{Martinson:2002}
\begin{subequations}
\begin{equation}
u(\zeta )=u_{2}+(u_{1}-u_{2})cn^{2}(\zeta \sqrt{\frac{u_{1}-u_{3}}{12\beta }}%
,q)\text{ \ , \ \ }q^{2}=\frac{u_{1}-u_{2}}{u_{1}-u_{3}}\text{ \ \ \ .}
\label{B12}
\end{equation}%
The function $w=cn(z,q)$ is the Jacobi elliptic function (elliptic cosine)
and it is defined as the inverse function to the elliptic integral of the
first kind
\end{subequations}
\begin{equation}
z=\int\limits_{0}^{w}\frac{d\varphi }{\sqrt{1-q^{2}\sin ^{2}\varphi }}\text{
\ \ . }  \label{B13}
\end{equation}%
The parameter $q$ in Eq. (\ref{B13}) and Eq. (\ref{B14}) is the modulus
parameter of the elliptic integral. For physical applications $0<q<1$, but
from a mathematical point of view it can also be imaginary. If $w=\frac{\pi
}{2}$ and $y=\sin \varphi $, then
\begin{equation}
K(q)\equiv \int\limits_{0}^{\frac{\pi }{2}}\frac{d\varphi }{\sqrt{%
1-q^{2}\sin ^{2}\varphi }}=\int\limits_{0}^{1}\frac{dy}{\sqrt{%
(1-y^{2})(1-q^{2}y^{2})}}\text{ \ \ \ }  \label{B14}
\end{equation}%
is called the elliptic integral of the first kind. The second integral in
Eq. (\ref{B14}) is written in the s.c. Legendre form. Since the solution Eq.
(\ref{B12}) physically describes the s.c. "knoidal wave", the elliptic
integral $K(q)$ is related to the space period $\lambda $ by the formulae
\begin{equation}
\lambda =4K(q)\sqrt{\frac{3\beta }{u_{1}-u_{3}}}\text{ \ \ \ \ .}
\label{B15}
\end{equation}%
In this paper we shall present an analytical approach for solving integrals
of the type Eq. (\ref{B14}) (however, the upper integration variable will
not be $1$, but will be arbitrary). This would mean also that in the Jacobi elliptic
function $cn(z,q)$ due to the defining equality Eq. (\ref{B13}), $z$ will
depend in a complicated way on the modulus parameter $q$. The Jacobi
elliptic sine function $y=sn(\sqrt{\frac{g}{a}}\frac{t-t_{0}}{q},q)$ had
been used also in the monograph by Whittaker \cite{Whittaker:1947} to denote
the solutions of the differential equation $\overset{.}{y}^{2}=\frac{g}{%
aq^{2}}(1-y^{2})(1-q^{2}y^{2})$, describing the motion of the mathematical
pendulum.

\section{\label{sec:DefinEllipticIntegrals}Standard mathematical definition
for elliptic integrals of higher order in the general case}

According to the general definition \cite{PrasSol:1997}, elliptic integrals
are of the type
\begin{equation}
\int R(y,\sqrt{G(y)})dy\text{ \ or}\int R(x,\sqrt{P(x)})dx\text{ \ \ ,}
\label{B16}
\end{equation}%
where $R(\widetilde{y},y)$ and $R(\widetilde{x},x)$ are rational functions
of the variables $\widetilde{y},y$ or $\widetilde{x},x$ and $\widetilde{y}=%
\sqrt{G(y)}$, $\widetilde{x}=\sqrt{P(x)}$ are arbitrary polynomials of the
fourth or of the third degree respectively. In accord with the notations in
the monograph \cite{PrasSol:1997}, the fourth-order and the third-order
polynomials will be of the form
\begin{equation}
\widetilde{y}^{2}=G(y)=a_{0}y^{4}+4a_{1}y^{3}+6a_{2}y^{2}+4a_{3}y+a_{4}\text{
\ \ , \ }\widetilde{x}^{2}=P(x)=ax^{3}+bx^{2}+cx+d\text{ \ .\ }  \label{B17}
\end{equation}%
So in the general case, elliptic integrals of the $n-$th order and of the
first kind are defined as
\begin{equation}
J_{n}^{(4)}=\int \frac{y^{n}dy}{\sqrt{%
a_{0}y^{4}+4a_{1}y^{3}+6a_{2}y^{2}+4a_{3}y+a_{4}}}\text{ \ , \ }%
J_{n}^{(3)}=\int \frac{x^{n}dx}{\sqrt{ax^{3}+bx^{2}+cx+d\text{ }}}
\label{B18}
\end{equation}%
Elliptic integrals of the $n-$th order and of the third kind are defined as
\begin{equation}
H_{n}^{(4)}=\int \frac{dy}{(y-c_{1})^{n}\sqrt{%
a_{0}y^{4}+4a_{1}y^{3}+6a_{2}y^{2}+4a_{3}y+a_{4}}}\text{ \ , \ }%
H_{n}^{(3)}=\int \frac{dx}{(x-c_{2})^{n}\sqrt{ax^{3}+bx^{2}+cx+d\text{ }}}%
\text{ \ .}  \label{B19}
\end{equation}%
In the above definitions, the upper indices "$(3)$" or "$(4)$" denote the
order of the algebraic polynomial under the square root in the denominator,
while the lower indice "$n$" denotes the order of the elliptic integral,
related to the "$y^{n}$" or "$x^{n}$" terms in the nominator of Eq. (\ref%
{B18}) and in the denominator of Eq. (\ref{B19}) (for $n$-th order elliptic
integrals of the fourth and of the third kind respectively). The contemporary definition of elliptic
integrals in \cite{PrasSol:1997} is fully consistent with the definition in
older monographs, such as \cite{Legendre:1825}. The elliptic integrals Eq. (%
\ref{B16}) - Eq. (\ref{B18}) can be considered to be a generalization of the
s.c. abelian integrals, defined in the monograph \cite{Fichten2} as
\begin{equation}
\int R(y,\sqrt[m]{\frac{\alpha y+\beta }{\gamma y+\delta }})dy\text{ \ \ ,
\ \ \ \ }\int R(y,\sqrt{ay^{2}+by+c}dy\text{ \ \ \ .}  \label{B20}
\end{equation}%
These integrals in \cite{Fichten2} have been treated analytically, the
second integral for example by means of the known Euler substitutions.

The elliptic integrals Eq. (\ref{B16}) - Eq. (\ref{B19}), however, are
believed to be not expressible by elementary functions. For example, in the
textbook \cite{Liashko:1983} it has been declared (although without any
proof) that "higher-order elliptic integrals in the Legendre form can be
expressed by means of the three standard integrals
\begin{equation}
\int \frac{dy}{\sqrt{(1-y^{2})(1-q^{2}y^{2})}}\text{ \ \ , \ \ }\int \frac{%
y^{2}dy}{\sqrt{(1-y^{2})(1-q^{2}y^{2})}}\text{\ \ , \ }\int \frac{dy}{%
(1+hy^{2})\sqrt{(1-y^{2})(1-q^{2}y^{2})}},  \label{B21}
\end{equation}%
where the parameter $h$ can be also a complex number". \ This assertion is
not precise, because in the paper \cite{DimitrovAIP:2022} an example was
found, when the second integral in Eq. (\ref{B21}) in fact can be expressed
through a logarithmic term. A similar statement to the one in \ \cite%
{Liashko:1983} is given also in the monograph \cite{Fichten2}: "Integrals of
the type Eq. (\ref{B16}) $\int R(y,\sqrt{G(y)})dy$ \ and $\ \int R(x,\sqrt{%
P(x)})dx$\ \ cannot be expressed by elementary functions in their final
form."

\section{ \label{sec:LegendreWeierstrass}Transforming an elliptic integral
in the Legendre form into an elliptic integral in the Weierstrass form}

The purpose of this section will be to propose a transformation, which will
transform the elliptic integral
\begin{equation}
\widetilde{J}_{0}^{(4)}(y)=\int \frac{dy}{\sqrt{(1-y^{2})(1-q^{2}y^{2})}}
\label{F0}
\end{equation}%
in the Legendre form into an elliptic integral in the Weierstrass form
\begin{equation}
\widetilde{J}_{0}^{(3)}(x)=-\sqrt{a}\int \frac{dx}{\sqrt{4x^{3}-g_{2}x-g_{3}}%
}\text{ \ \ \ .}  \label{F1}
\end{equation}%
For the purpose, the transformation
\begin{equation}
x=\frac{a}{y^{2}}+b  \label{F2}
\end{equation}%
shall be applied. It is possible to calculate the parameters $a$ and $b$, so
that the integral Eq. (\ref{F1}) is satisfied
\begin{equation}
a=\frac{9g_{3}}{g_{2}}K(q)\text{ \ \ \ ,}  \label{F3}
\end{equation}%
\begin{equation}
b=-\frac{a}{3}(1+q^{2})=-\frac{3g_{3}}{g_{2}}K(q)\text{ \ \ \ \ }  \label{F4}
\end{equation}%
and $K(q)$ is a rational function, depending on the modulus parameter $q$ of
the elliptic integral
\begin{equation}
K(q)\equiv \frac{q^{4}-q^{2}+1}{2q^{4}-5q^{2}+2}\text{ \ \ .}  \label{F5}
\end{equation}%
However, $g_{2}$ and $g_{3}$ remain undetermined. This will turn out to be
an advantage of the applied formalism, since further another representation of the integral
in the Weierstrass form shall be found with different Weierstrass invariants $%
\overline{g}_{2}$ and $\overline{g}_{3}$, which will depend explicitly on
the modulus parameter $q$. Then by comparing both representations in the
Weierstrass form and proving a new theorem, $g_{2}$ and $g_{3}$ will be
expressed through $\overline{g}_{2}$ and $\overline{g}_{3}$ in a complicated
way.

Further the following important property will be very important.

\bigskip Corollary: The parameter $b$ satisfies the right-hand side of the
parametrizable form  of the cubic equation, i.e.
\begin{equation}
4b^{3}-g_{2}b-g_{3}=0\text{ \ . }  \label{F6}
\end{equation}

The proof is straightforward, if the expressions for $g_{2}$ and $g_{3}$ are
expressed from Eq. (\ref{F3}) and Eq. (\ref{F4}) and are substituted into
the above cubic equation.

\section{ \label{sec:IntegralRecurrent}Equivalence between a formulae from
integral calculus and the recurrent system of equations for the elliptic
integral in terms of the $y-$variable}

Due to the property Eq. (\ref{F6}), the elliptic integral Eq. (\ref{F1}) can
be represented in the equivalent way
\begin{equation}
\widetilde{J}_{0}^{(3)}\text{ }=-\frac{\sqrt{a}}{2}\int \frac{d\overline{x}}{%
\overline{x}^{\frac{1}{2}}\sqrt{\overline{x}^{2}+3b\overline{x}+(b^{2}-\frac{%
g_{2}}{4})}}\text{ \ \ \ ,}  \label{F7}
\end{equation}%
where $\overline{x}\equiv x-b$. Such integrals in the monograph by Timofeev
\cite{Timofeev:1948} are analytically computed
\begin{equation*}
\int \frac{d\overline{x}}{\overline{x}^{m}\sqrt{\left( \widetilde{a}+2%
\widetilde{b}\overline{x}+\widetilde{c}\overline{x}^{2}\right) ^{2n+1}}}=-%
\frac{\sqrt{\left( \widetilde{a}+2\widetilde{b}\overline{x}+\widetilde{c}%
\overline{x}^{2}\right) ^{2n-1}}}{(m-1)\widetilde{a}\overline{x}^{m-1}}
\end{equation*}%
\begin{equation*}
-\frac{(2m+2n-3)\widetilde{b}}{(m-1)\widetilde{a}}\int \frac{d\overline{x}}{%
\overline{x}^{m-1}\sqrt{\left( \widetilde{a}+2\widetilde{b}\overline{x}+%
\widetilde{c}\overline{x}^{2}\right) ^{2n+1}}}
\end{equation*}%
\begin{equation}
-\frac{(m+2n-2)\widetilde{c}}{(m-1)\widetilde{a}}\int \frac{d\overline{x}}{%
\overline{x}^{m-2}\sqrt{\left( \widetilde{a}+2\widetilde{b}\overline{x}+%
\widetilde{c}\overline{x}^{2}\right) ^{2n+1}}}\text{ \ \ .}  \label{F8}
\end{equation}%
Since this is a formulae from integral calculus, it is interesting to prove
the following theorem, which most surprisingly establishes the equivalence
between the above relation from integral calculus with some of the recurrent
system of equations for the case of a fourth-degree polynomial in the
denominator of the elliptic integral \cite{PrasSol:1997}. The recurrent
system for the case of a cubic polynomial under the square root has been
investigated in the monographs of Fichtenholz \cite{Fichten2} and Smirnov
\cite{Smirnov3}. Below the formulation of the newly proved theorem is given.

\ Theorem: The integral Eq. (\ref{F8}) is equivalent to the second equation
\begin{equation}
\overline{z}^{.}\widetilde{Z}=3\widetilde{c}J_{4}^{(4)}(\overline{z})+4%
\widetilde{b}\text{ }J_{2}^{(4)}(\overline{z})+\widetilde{a}J_{0}^{(4)}(%
\overline{z})\text{ \ \ \ .}  \label{A57}
\end{equation}%
from the recurrent system of equations
\begin{equation}
\overline{z}^{m}\widetilde{Z}=(m+2)\widetilde{c}J_{m+3}^{(4)}(\overline{z}%
)+(m+1)2\widetilde{b}\text{ }J_{m+1}^{(4)}(\overline{z})+m\widetilde{a}%
J_{m-1}^{(4)}(\overline{z})\text{ \ \ ,}  \label{A55}
\end{equation}%
where  the fourth-degree polynomial $\widetilde{Z}$ is defined as
\begin{equation}
\widetilde{Z}^{2}=a_{0}\overline{z}^{4}+4a_{1}\overline{z}^{3}+6a_{2}%
\overline{z}^{2}+4a_{3}\overline{z}+a_{4}\text{ \ \ .}  \label{A50}
\end{equation}%
The  coefficients in the above polynomial have the values
\begin{equation}
a_{0}=\widetilde{c}\text{ \ \ , \ }a_{1}=0\text{ \ , \ \ }6a_{2}=2\widetilde{%
b}\text{ \ , \ }a_{3}=0\text{ \ \ , \ \ }a_{4}=\widetilde{a}\text{ \ \ . }
\label{A51}
\end{equation}%
The corresponding elliptic integrals of the first kind and of the $n-$th
order are
\begin{equation}
J_{n}^{(4)}(\overline{z})=\int \frac{\overline{z}^{n}}{\sqrt{\widetilde{a}+2%
\widetilde{b}\overline{z}^{2}+\widetilde{c}\overline{z}^{4.}}}d\overline{z}%
\text{ }=\int \frac{\overline{z}^{n}d\overline{z}}{\sqrt[.]{\overline{z}%
^{4}+3b\overline{z}^{2}+(3b^{2}-\frac{g_{2}}{4})}}\text{ \ \ \ ,}
\label{A53}
\end{equation}%
\begin{equation}
\text{ \ \ }H_{n}^{(4)}(\overline{z})=\int \frac{d\overline{z}}{\overline{z}%
^{n}\sqrt{\widetilde{a}+2\widetilde{b}\overline{z}^{2}+\widetilde{c}%
\overline{z}^{4.}}}=\text{ }\int \frac{d\overline{z}}{\overline{z}^{n}\sqrt[.%
]{\overline{z}^{4}+3b\overline{z}^{2}+(3b^{2}-\frac{g_{2}}{4})}}\text{\ \ \ .%
}  \label{A54}
\end{equation}

\bigskip Due to the established equivalence, it can be concluded that the
integral Eq. (\ref{F8}) cannot be used for the calculation of the zero-order
elliptic integral in the Legendre form Eq. (\ref{F0}), because the recurrent
relation Eq. (\ref{A57}) contains also the higher-order elliptic integrals $%
J_{2}^{(4)}(\overline{z})$ and $J_{4}^{(4)}(\overline{z})$.

\section{ \label{sec:TheoremRecurrent}A new theorem, concerning the
equivalence between some of the equations in the recurrent system}

Theorem: From Eq. (\ref{A55}),  the corresponding equation for $m=-3$ \
\begin{equation}
\frac{\widetilde{y}}{y^{3}}%
=-q^{2}J_{0}(y)+2(1+q^{2})H_{2}^{(4)}(y)-3H_{4}^{(4)}(y)  \label{F9}
\end{equation}%
can be transformed into the equation for $m=1$
\begin{equation}
y\widetilde{y}^{(q)}=J_{0}(y)-2(1+q^{2})J_{2}(y)+3q^{2}J_{4}(y)  \label{F10}
\end{equation}%
by means of the transformation $y\Rightarrow \frac{1}{qy}$ and the simple
relations (for the concrete case, for $m=2$)
\begin{equation}
H_{m}(y)=-q^{m}J_{m}(\frac{1}{qy})\text{ \ \ \ \ , \ \ }H_{m}(\frac{1}{qy}%
)=-q^{m}J_{m}(y)  \label{F11}
\end{equation}%
In such a way, only one independent equation can be obtained from the $4D$ recurrent system
of equations

\begin{equation}
2(1+q^{2})J_{2}^{(q)}(y)=3q^{2}J_{4}^{(q)}(y)+J_{0}^{(q)}(y)-y\widetilde{y}%
^{(q)}\text{ \ \ \ .}  \label{F11A1}
\end{equation}

\section{\protect\bigskip \label{sec:TwoRecurSystems}Relations between the
recurrent system of equations in terms of the $y$ and $x$ variables}

\bigskip

\subsection{\protect\bigskip \label{sec:RecurXVariable}The recurrent system
of equations in terms of the $x$ variable}

Now we shall write the recurrent system of equations for the elliptic
integrals

\begin{equation}
I_{m}^{(3)}\equiv \int \frac{x^{m}dx}{\sqrt{4x^{3}-g_{2}x-g_{3}}}\text{ \ \
\ \ , \ \ \ \ \ \ \ \ \ \ }G_{m}^{(3)}\equiv \int \frac{dx}{x^{m}\sqrt{%
4x^{3}-g_{2}x-g_{3}}}\text{ \ \ , \ \ \ \ }x=\frac{\widetilde{a}_{2}}{y^{2}}+%
\widetilde{b}_{2}\text{ \ \ ,}  \label{F12}
\end{equation}%
as this has been performed in the monographs \ \cite{Fichten2} and \cite%
{Smirnov3}. Generally, the polynomial of the third-order under the square
root is denoted as
\begin{equation}
X^{2}=a_{0}x^{4}+4a_{1}x^{3}+6a_{2}x^{2}+4a_{3}x+a_{4}\text{ \ \ ,}
\label{F13}
\end{equation}%
but for the concrete case of the integrals Eq. (\ref{F12}) \ the coefficients
will be given by
\begin{equation}
a_{0}=0\text{ \ \ , \ }a_{1}=1\text{ \ , \ \ }a_{2}=0\text{ \ , \ }a_{3}=-%
\frac{g_{2}}{4}\text{ \ \ , \ \ }a_{4}=-g_{3}\text{ \ \ . }  \label{F14}
\end{equation}%
The recurrent system of equations is easily found to be
\begin{equation}
x^{n}X=2(2n+3)I_{n+2}-\frac{g_{2}}{2}(2n+1)I_{n}-ng_{3}I_{n-1}\text{ \ \ .}
\label{F15}
\end{equation}%
It should be noted also that the variable transformation $x=\frac{\widetilde{%
a}_{2}}{y^{2}}+\widetilde{b}_{2}$ in Eq. (\ref{F12}) contains another
parameters $\widetilde{a}_{2}$ and $\widetilde{b}_{2}$, which are different
from the parameters $a_{2}$ and $b_{2}$ in the transformation Eq. (\ref{F2}%
). Now we shall transform the three-degree polynomials under the
square root in Eq. (\ref{F12}) into the known four-degree polynomials for
the elliptic integral in the Legendre form. The transformation \ $x=\frac{%
\widetilde{a}_{2}}{y^{2}}+\widetilde{b}_{2}$ (as well as the previously
applied transformation (\ref{F2}) $x=\frac{a}{y^{2}}+b$) are
generalizations of the transformation $x=e_{3}+\frac{(e_{1}-e_{3})}{y^{2}}$
in the paper \cite{Kraniotis:2004}, where $e_{1},e_{2},e_{3}$ are constant
roots of a cubic polynomial. The transformation transforms the integral $%
\int \frac{dx}{\sqrt{4(x-e_{1})(x-e_{2})(x-e_{3})}}$ into the integral $-%
\frac{1}{\sqrt{e_{1}-e_{3}}}\int \frac{dy}{\sqrt{(1-y^{2})(1-q^{2}y^{2})}}$%
, where $q^{2}=\frac{(e_{2}-e_{3})}{(e_{1}-e_{3})}$. Under another
transformation $y\Rightarrow \frac{1}{\widetilde{k}y}$, the $x-$variable
will transform as

\begin{equation}
\text{ \ \ \ \ \ \ \ }\widetilde{x}\equiv \widetilde{a}_{2}\widetilde{k}%
^{2}y^{2}+\widetilde{b}_{2}\text{\ \ \ \ \ ,}  \label{F16}
\end{equation}%
where a new variable $\widetilde{x}$ has been introduced. The two recurrent
system of equations Eq. (\ref{A55}) (written in terms of the $y-$variable)
and the recurrent system Eq. (\ref{F15}) for the $x-$variable are not
independent because
\begin{equation}
I_{0}^{(3)}(x)=-\frac{J_{0}^{(q)}(y)}{\left( \sqrt{a}\right) ^{q}}=\frac{2%
\widetilde{a}_{2}}{\left( \sqrt{a}\right) ^{q}}J_{0}^{(\widetilde{k})}(y)%
\text{ \ \ \ ,}  \label{F17}
\end{equation}%
where $\widetilde{k}$ is determined so that $4x^{3}-g_{2}(q)x-g_{3}(q)$ is
transformed into the expression $(1-y^{2})(1-$ $\widetilde{k}^{2}y^{2})$.
Consequently, the last expression is satisfied if $\widetilde{k}$ is defined
as
\begin{equation}
\widetilde{k}=-1-\frac{6\alpha }{\sqrt[3]{2}}\text{ \ , \ }\alpha =%
\widetilde{b}_{2}\text{ is a root of \ }4\widetilde{b}_{2}^{3}-g_{2}^{(q)}%
\widetilde{b}_{2}-g_{3}^{(q)}=0\text{ \ \ , \ \ }\widetilde{a}_{2}=\frac{1}{%
\sqrt[3]{4}}=\frac{\widetilde{k}^{4}-\widetilde{k}^{2}+1}{3g_{2}^{(q)}}\text{%
\ }\ \ \ \ \ \text{. }  \label{F18}
\end{equation}%
Two central problems remain to be solved:

1. How to combine the derived equations in terms of the $x$ variable with
the found relation Eq. (\ref{F11A1})?

2. How to estimate the parameter $g_{2}$ in Eq. (\ref{F18})? In fact,
further in the next sections it will be proved that $g_{2}=g_{2}^{(q)}\equiv
g_{2}(q)$ and $g_{3}=g_{3}^{(q)}\equiv g_{3}(q)$\ will depend in a
complicated way on the modulus parameter $q$, but this dependence will be
found from another Weierstrass representation of the integral $%
J_{0}^{(4)}(y) $ Eq. (\ref{F0}). This new representation will not contain
the conformal factor $\sqrt{a}$ in Eq. (\ref{F1}), but instead will be
characterized by different Weierstrass invariants $\overline{g}_{2}^{(q)}$
and $\overline{g}_{3}^{(q)}$, which will be explicitly determined by the
method of the s.c. "fourth-degree polynomial uniformization". This algebraic method will be
considered in details in the following sections.

\subsection{\label{sec:CombinTwoRecur}Combining the two recurrent system of
equations}

From the two recurrent system of equations in terms of the $y-$variable and
the $x-$variable, the following equation can be obtained

\begin{equation}
\widetilde{X}^{(\widetilde{k})}=N_{1}(\widetilde{a}_{2}\text{,}\widetilde{b}%
_{2}\text{,}\widetilde{k})J_{2}^{(\widetilde{k})}(y)+N_{2}(\widetilde{a}_{2}%
\text{,}\widetilde{b}_{2}\text{,}\widetilde{k})J_{0}^{(\widetilde{k}%
)}(y)+N_{3}(\widetilde{a}_{2}\text{,}\widetilde{b}_{2}\text{,}\widetilde{k})=
\label{F19}
\end{equation}%
\begin{equation}
=N_{1}^{(\widetilde{k})}J_{2}^{(\widetilde{k})}(y)+N_{2}^{(\widetilde{k})}%
\left[ -\frac{1}{2\widetilde{a}_{2}}J_{0}^{(q)}(y)\right] +N_{3}^{(%
\widetilde{k})}\text{ \ \ \ ,}  \label{F19A}
\end{equation}%
where
\begin{equation}
N_{1}(\widetilde{a}_{2}\text{,}\widetilde{b}_{2}\text{,}\widetilde{k})\equiv
N_{1}^{(\widetilde{k})}\equiv 8\widetilde{a}_{2}^{3}\widetilde{k}^{2}(1+%
\widetilde{k}^{2})+24\widetilde{a}_{2}^{2}\widetilde{b}_{2}\widetilde{k}^{2}%
\text{ \ \ ,}  \label{F20}
\end{equation}%
\begin{equation}
N_{2}(\widetilde{a}_{2}\text{,}\widetilde{b}_{2}\text{,}\widetilde{k}%
,)\equiv N_{2}^{(\widetilde{k})}\equiv 12\widetilde{a}_{2}\widetilde{b}%
_{2}^{2}-\frac{g_{2}\widetilde{a}_{2}}{(\sqrt{a})^{(q)}}-4\widetilde{a}%
_{2}^{3}\widetilde{k}^{2}\text{ \ \ \ ,}  \label{F21}
\end{equation}%
\begin{equation}
N_{3}(\widetilde{a}_{2}\text{,}\widetilde{b}_{2}\text{,}\widetilde{k})\equiv
N_{3}^{(\widetilde{k})}\equiv 4\widetilde{a}_{2}^{3}\widetilde{k}^{2}y^{2}%
\widetilde{y}^{(\widetilde{k})}\text{ \ \ \ \ }  \label{F22}
\end{equation}%
and \ $\widetilde{X}^{(\widetilde{k})}\equiv 4\widetilde{x}^{3}-g_{2}^{(%
\widetilde{k})}\widetilde{x}-g_{3}^{(\widetilde{k})}$. The notations $%
g_{2}^{(\widetilde{k})}$ and $g_{3}^{(\widetilde{k})}$ mean that these
Weierstrass invariants are expressed as complicated functions of the
parameter $\widetilde{k}$.

Now it can be noted that if it is possible to make the replacement $%
\widetilde{k}\Longleftrightarrow q$, then another equation will be obtained
\begin{equation}
\widetilde{X}%
^{(q)}=N_{1}^{(q)}J_{2}^{(q)}(y)+N_{2}^{(q)}J_{0}^{(q)}(y)+N_{3}^{(q)}\text{
\ \ \ }  \label{F23}
\end{equation}%
and in this way, the equations Eq. (\ref{F19A}) and Eq. (\ref{F23}) will
give the following solution for the zero-order elliptic integral in the
Legendre form \ \
\begin{equation}
J_{0}^{(4)}(y)\equiv J_{0}^{(q)}(y)=\frac{2\widetilde{X}^{(q)}}{\sqrt[3]{2}%
N_{2}^{(q)}}-\frac{N_{3}^{(q)}}{N_{2}^{(q)}}-\frac{\sqrt[3]{2}\widetilde{X}%
^{(\widetilde{k})}\left( \sqrt[3]{2}-2N_{2}^{(q)}\right) }{N_{2}^{(q)}\left(
\sqrt[3]{2}-2\right) }-\frac{N_{1}^{(q)}\left( 2\widetilde{X}^{(q)}-\sqrt[3]{%
2}N_{3}^{(q)}\right) }{N_{2}^{(q)}\sqrt[3]{2}}\text{ \ \ \ .}  \label{F24}
\end{equation}%
In view of the replacement $\widetilde{k}\Longleftrightarrow q$, one also
has the following equality, which follows from the defining relations Eq. (%
\ref{F18})
\begin{equation}
\frac{\widetilde{k}^{4}-\widetilde{k}^{2}+1}{3g_{2}(\widetilde{k})}=\frac{%
q^{4}-q^{2}+1}{3g_{2}(q)}\text{ \ \ \ \ \ .}  \label{F25}
\end{equation}%
If this is a quite a general condition (not imposing any restrictions), \
then the replacement $\widetilde{k}\Longleftrightarrow q$ will be justified.
Evidently this will depend on the determination of $g_{2}^{(q)}\equiv
g_{2}(q)$, which will be given in the next sections.

\section{\label{sec:4D uniformization}Another integral in the Weierstrass
form after applying the  four-dimensional uniformization}

Further by "four-dimensional uniformization" it shall be meant that the
polynomial of the fourth degree under the square root in the denominator of
the elliptic integral shall be "uniformized" by introducing a complicated
dependence on the Weierstrass function and its derivative, which represent
functions of a complex variable. The term "uniformization" shall be clarified in the
next section.

\subsection{\label{sec:3D uniformization}The standard method for $3D$ uniformization in
elliptic functions theory-reminder of the basic facts}

Let in accord with the standard notions in elliptic functions theory \cite%
{PrasSol:1997} and \cite{Husemoller:1987}, the s.c. " fundamental
parallelogram" is defined on the two-periodic lattice \
\begin{equation}
\Pi \equiv \left\{ m\omega _{1}+n\omega _{2}\text{ ; \ \ \ }0\leq m\leq 1%
\text{ , }0\leq n\leq 1\text{ , Imag}\left( \frac{\omega _{1}}{\omega _{2}}%
\right) >0\right\} \text{ \ .}  \tag{II.9}
\end{equation}%
The two periods $\omega _{1}$ and $\omega _{2}$ constitute the basis on the
complex plane and thus each complex number can be represented as $z=$ $m\omega
_{1}+n\omega _{2}$. Then the Weierstrass function
\begin{equation}
\rho (z)\equiv \frac{1}{z^{2}}+\sum \left( \frac{1}{(z-\omega )^{2}}-\frac{1%
}{\omega ^{2}}\right)  \label{G1}
\end{equation}%
and its first derivative $\rho ^{^{\prime }}(z)\equiv -\frac{2}{z^{3}}%
-2\sum \frac{1}{(z-\omega )^{3}}$ satisfy the parametrizable form of the
cubic algebraic equation
\begin{equation}
y^{2}=4x^{3}-g_{2}x-g_{3}\text{ \ \ i.e. \ \ \ \ }x\equiv \rho (z)\text{ \ \
\ , \ \ }y\equiv \rho ^{^{\prime }}(z)\text{ \ \ \ \ , }  \label{G2}
\end{equation}%
$g_{2}$ and $g_{3}$ are the s.c. "invariants of the Weierstrass function" $%
\rho (z)$
\begin{equation}
g_{2}=60G_{4}=60\sum \frac{^{\prime }1}{\omega ^{4}}\text{ \ , \ }%
g_{3}=140G_{6}=140\sum \frac{^{\prime }1}{\omega ^{6}}\text{ \ \ ,}
\tag{II.8}
\end{equation}%
defined as infinite convergent sums over the period $\omega $ of the
two-periodic lattice. The prime "$^{\prime }$" above the two sums means that
the period $\omega =0$ is excluded from the summation, so that the
expressions would not tend to infinity.

\subsection{\label{sec:4D uniformizationRevNew}Uniformization of
four-dimensional algebraic equations and application to the theory of
elliptic integrals. Another representation of the integral in the Weierstrass
form}

\bigskip The derivation of the second representation of the elliptic
integral in the Weierstrass form is based on the following theorem, proved
in the monograph by Whittaker and Watson \cite{WhittakerWatson:1927}.

Theorem: Let
\begin{equation}
J_{0}^{(4)}(y)=\int \frac{dy}{\sqrt{%
a_{0}y^{4}+4a_{1}y^{3}+6a_{2}y^{2}+4a_{3}y+a_{4}}}\text{ \ \ }  \tag{II.11A}
\end{equation}%
is a zero-order elliptic integral with the four-dimensional polynomial under
the square root in the denominator
\begin{equation}
Y\equiv f(y)\equiv a_{0}y^{4}+4a_{1}y^{3}+6a_{2}y^{2}+4a_{3}y+a_{4}\text{ }\
\ \text{, \ \ \ }f(y_{0})=0\text{\ \ \ .}  \label{C1}
\end{equation}%
Then after the variable changes
\begin{equation}
\frac{1}{t-y_{0}}=\frac{\sigma -\frac{1}{2}A_{2}}{A_{3}}\text{ \ \ , \ \ \ \
}\frac{1}{y-y_{0}}=\frac{s-\frac{1}{2}A_{2}}{A_{3}}\text{\ }  \label{C2}
\end{equation}%
the integral can be rewritten as
\begin{equation}
I_{0}^{(3)}(x)=\int_{s}^{\infty }\frac{d\sigma }{\sqrt{4\sigma ^{3}-%
\overline{g}_{2}\sigma -\overline{g}_{3}}}\text{ }\ \ \ \text{,\ \ }
\label{C3}
\end{equation}%
where the Weierstrass invariants $\overline{g}_{2}$ and $\overline{g}_{3}$
can be exactly calculated \ as
\begin{equation}
\overline{g}_{2}=3A_{2}^{2}-4A_{1}A_{3}\text{ \ \ \ \ ; \ \ }\overline{g}%
_{3}=2A_{1}A_{2}A_{3}-A_{2}^{3}-A_{0}A_{3}^{2}  \label{C4}
\end{equation}%
and $A_{0},A_{1},A_{2},A_{3}$ are the introduced notations for the
polynomial expressions
\begin{equation}
A_{0}\equiv a_{0}\text{ \ \ , \ \ }A_{1}\equiv a_{0}y_{0}+a_{1}\text{ \ \ \
\ ,}  \label{C5}
\end{equation}%
\begin{equation}
A_{2}\equiv a_{0}y_{0}^{2}+2a_{1}y_{0}+a_{2}\text{ \ \ , \ \ \ }A_{3}\equiv
a_{0}y_{0}^{3}+3a_{1}y_{0}^{2}+3a_{2}y_{0}+a_{3}\text{ \ \ \ \ .}  \label{C6}
\end{equation}

In the concrete case for the polynomial in the Legendre form%
\begin{equation}
f(y)\equiv (1-y^{2})(1-q^{2}y^{2})=1-(1+q^{2})y^{2}+y^{4}\text{ \ \ , }
\label{C7}
\end{equation}%
one can compute $\overline{g}_{2}^{(1,2)}$ and $\overline{g}_{3}^{(1,2)}$
(the roots of the polynomial $f(y)$ are $y_{1,2}=\pm 1$, \ $y_{3,4}=\pm
\frac{1}{q}$)
\begin{equation}
\overline{g}_{2}^{(1,2)}\equiv q^{2}+\frac{1}{12}(1+q^{2})^{2}>0  \label{C8}
\end{equation}%
\begin{equation}
\overline{g}_{3}^{(1,2)}\equiv \frac{q^{6}}{4}-\frac{q^{4}}{6}-\frac{5q^{2}}{%
12}-\frac{1}{6^{3}}\left[ 5q^{2}-1\right] ^{3}\text{\ \ }  \label{C9}
\end{equation}%
\begin{equation}
=\left( \frac{1}{4}-\frac{5^{3}}{6^{3}}\right) q^{6}+\left( -\frac{1}{6}+%
\frac{75}{6^{3}}\right) q^{4}-\left( \frac{1}{12}+\frac{15}{6^{3}}\right)
q^{2}+\frac{1}{6^{3}}\text{ \ \ \ .}  \label{C10}
\end{equation}%
Similar expressions for $\overline{g}_{2}^{(3,4)}$ and $\overline{g}%
_{3}^{(3,4)}$ can also be written, but they will not be given here. Here 
the important fact, related to the theory of elliptic functions is that if $%
\overline{g}_{2}^{(1,2)}$, $\overline{g}_{3}^{(1,2)}$ or $\overline{g}%
_{2}^{(3,4)}$, $\overline{g}_{3}^{(3,4)}$ are known, then the Weierstrass
function $\rho (z;\overline{g}_{2},\overline{g}_{3})$ is uniquely
determined. \

\bigskip Theorem: (another formulation) If the Weierstrass elliptic function is defined as $%
\rho (z;\overline{g}_{2},\overline{g}_{3})$, then the two variables
\begin{equation}
y=y_{0}+\frac{1}{4}\frac{f^{%
%TCIMACRO{\U{b4}}%
%BeginExpansion
{\acute{}}%
%EndExpansion
}(y_{0})}{\left[ \rho (z;\overline{g}_{2},\overline{g}_{3})-\frac{1}{24}f^{%
%TCIMACRO{\U{b4}}%
%BeginExpansion
{\acute{}}%
%EndExpansion
%TCIMACRO{\U{b4}}%
%BeginExpansion
{\acute{}}%
%EndExpansion
}(y_{0})\right] }\text{ \ \ \ ,}  \label{C13}
\end{equation}%
\begin{equation}
Y=-\frac{1}{4}\frac{f^{%
%TCIMACRO{\U{b4}}%
%BeginExpansion
{\acute{}}%
%EndExpansion
}(y_{0})\rho ^{%
%TCIMACRO{\U{b4}}%
%BeginExpansion
{\acute{}}%
%EndExpansion
}(z;\overline{g}_{2},\overline{g}_{3})}{\left[ \rho (z;\overline{g}_{2},%
\overline{g}_{3})-\frac{1}{24}f^{%
%TCIMACRO{\U{b4}}%
%BeginExpansion
{\acute{}}%
%EndExpansion
%TCIMACRO{\U{b4}}%
%BeginExpansion
{\acute{}}%
%EndExpansion
}(y_{0})\right] ^{2}}\text{ \ \ \ }  \label{C14}
\end{equation}%
define an unique parametrization of the $4D$ algebraic curve Eq. (\ref{C1}).

Consequently,  the method for  "four-dimensional parametrization" can be considered to be the analogue
of the known "three-dimensional parametrization", given by Eq. (\ref{G2}%
).

\subsection{\label{sec:SecMethod4D uniformization}Second method for
four-dimensional uniformization}

This method is also exposed in the monograph by E.T. Whittaker and G.
N.Watson \cite{WhittakerWatson:1927}. We shall review the corresponding
theorem by making a slight modification in the fourth-degree polynomial Eq. (%
\ref{C1})- the numbers $4,6$ and $4$ in the coefficients are removed.

\bigskip Theorem: If \ the variable transformation $z=\int\limits_{y_{0}}^{y}%
\frac{dt}{\sqrt{f(t)}}$ is performed ($%
f(t)=a_{0}y^{4}+a_{1}y^{3}+a_{2}y^{2}+a_{3}y+a_{4}$) in the $4D$ elliptic
integral

\begin{equation}
J_{0}^{(4)}(y)=\int \frac{dy}{\sqrt{%
a_{0}y^{4}+a_{1}y^{3}+a_{2}y^{2}+a_{3}y+a_{4}}}\text{ \ \ ,}  \label{II.11AB}
\end{equation}%
then the Weierstrass invariants $\overline{g}_{2}$ and $\overline{g}_{3}$ in
the integral (\ref{C3}) \ $I_{0}^{(3)}(x)=\int_{s}^{\infty }\frac{d\sigma }{%
\sqrt{4\sigma ^{3}-\overline{g}_{2}\sigma -\overline{g}_{3}}}$ are uniquely
calculated as
\begin{equation}
\overline{g}_{2}=a_{0}a_{4}-\frac{1}{4}a_{1}a_{3}+\frac{1}{12}a_{2}^{2}\text{
\ \ ,}  \label{C15}
\end{equation}%
\begin{equation}
\overline{g}_{3}=\frac{1}{6}a_{0}a_{2}a_{4}+\frac{1}{48}a_{1}a_{2}a_{3}-%
\frac{a_{2}^{3}}{6^{3}}-\frac{1}{16}(a_{0}a_{3}+a_{1}^{2}a_{4})\text{ \ .}
\label{C16}
\end{equation}

As an example, which will be used further, for the polynomial
\begin{equation}
f(y)\equiv (1-y^{2})(1-q^{2}y^{2})=1-(1+q^{2})y^{2}+y^{4}\text{ \ \ , }
\label{C17}
\end{equation}%
the two invariants are calculated to be
\begin{equation}
\overline{g}_{2}=q^{2}+\frac{1}{12}(1+q^{2})^{2}\text{ }>0\text{\ \ ,}
\label{C18}
\end{equation}%
\begin{equation}
\overline{g}_{3}=-\frac{1}{6}q^{2}(1+q^{2})+\frac{(1+q^{2})^{3}}{6^{3}}\text{
\ \ \ \ .}  \label{C19}
\end{equation}

Clearly, the result for $\overline{g}_{2}$ in Eq. (\ref{C18}) coincides with
the result for $\overline{g}_{2}^{(1,2)}\equiv q^{2}+\frac{1}{12}%
(1+q^{2})^{2}>0$ in Eq. (\ref{C8}), but formulae Eq. (\ref{C19}) for $%
\overline{g}_{3}$ is quite different from formulaes Eq. (\ref{C9}) and Eq. (%
\ref{C10}) for $\overline{g}_{3}^{(1,2)}$. The same refers also to the
formulae for $\overline{g}_{3}^{(3,4)}$, not presented here.

\bigskip Important consequences: 1. For this case there is only one pair of
Weierstrass invariants $\overline{g}_{2}$ and $\overline{g}_{3}$ instead of
the two pairs $\overline{g}_{2}^{(1,2)},\overline{g}_{3}^{(1,2)}$ and $%
\overline{g}_{2}^{(3,4)},\overline{g}_{3}^{(3,4)}$ for the previous case,
considered in the preceding section \ref{sec:4D uniformizationRevNew}.

2. The expression for $\overline{g}_{2}$ Eq. (\ref{C18}) coincides with the
expression for $\overline{g}_{2}^{(1,2)}$ in Eq. (\ref{C8}), but the
expression $\overline{g}_{3}$ Eq. (\ref{C19}) is different from the
expression Eq. (\ref{C9}) for $\overline{g}_{2}^{(1,2)}$. It is not clear
whether this will be so for the case of an elliptic integral with an
arbitrary polynomial in the denominator. In the next section, however, it
will be shown that if the four-dimensional integral Eq. (\ref{B5}) is taken,
then the Weierstrass invariants $\overline{g}_{2}$ and $\overline{g}_{3}$
will be given by one and the same expressions, if first calculated according
to formulaes Eq. (\ref{C5}) and Eq. (\ref{C6}) in the previous section, and afterwards -
according to the formulaes Eq. (\ref{C15}) and Eq. (\ref{C16}) in this section.

3. The two invariants $\overline{g}_{2}$ and $\overline{g}_{3}$ are
symmetrical with respect to the interchanges $a_{0}\rightleftarrows a_{4}$
and also $a_{1}\rightleftarrows a_{3}$. The polynomial $f(y)$ is also
invariant with respect to these changes.

\subsection{\label{sec:ExampleEllipticCosmol}An example with the elliptic
integral Eq. (\protect\ref{B5}) from the expression for the apparent
luminosity distance in cosmology}

The integral Eq. (\ref{B5}) is another example, when the two methods for
uniformization can be applied - the first method when the Weierstrass
invariants $\overline{g}_{2}$ and $\overline{g}_{3}$ are calculated
according to the formulaes Eq. (\ref{C4}) - Eq. (\ref{C7}) for the
coefficient functions $A_{0},A_{1},A_{2}$, and the second method- according
to formulaes Eq. (\ref{C15}) - Eq. (\ref{C16}) for the coefficients $%
a_{0},a_{1},a_{2},a_{3}$ in the fourth-degree polynomial under the integral
expression in the elliptic integral Eq. (\ref{II.11AB}). For this integral,
the coefficients are equal correspondingly to the following cosmological
energy densities
\begin{equation}
a_{0}\equiv \Omega _{\Lambda }\text{ \ \ , \ \ }a_{1}\equiv 0\text{ \ , \ }%
a_{2}\equiv \Omega _{K}\text{ \ , \ }a_{3}\equiv \Omega _{M}\text{ \ , \ }%
a_{4}\equiv \Omega _{R}\text{ \ \ .}  \label{C20}
\end{equation}%
The first method makes use of the s.c. "turning point $y_{0}$", which is a
root of the fourth-order equation $f(y_{0})=\Omega _{\Lambda }y_{0}^{4}+\Omega
_{K}y_{0}^{2}+\Omega _{M}y_{0}+\Omega _{R}=0$. Physically, the point $y_{0}$
is related to the possible change of the expansion of the Universe with
contraction. Remarkably, the calculation of the invariants $\overline{g}_{2}$
and $\overline{g}_{3}$ according to the two methods gives one and the same
result, unlike the previous case
\begin{equation}
\overline{g}_{2}=\Omega _{\Lambda }\Omega _{R}+\frac{1}{12}\Omega _{K}^{2}%
\text{ \ \ , \ \ }  \label{C21}
\end{equation}%
\begin{equation}
\overline{g}_{3}=\frac{1}{6}\Omega _{\Lambda }\Omega _{K}\Omega _{R}-\frac{%
\Omega _{K}^{3}}{6^{3}}-\Omega _{\Lambda }\frac{\Omega _{M}^{2}}{16}\text{ \
\ \ .}  \label{C22}
\end{equation}

\subsubsection{\protect\bigskip \label{sec:MasslessNeutrinos}The subcase of
massless neutrinos}

Now let us estimate numerically these expressions for the values $\Omega
_{\Lambda }=0.67$ and $\Omega _{M}=0.33$, taken from the monograph \cite%
{Weinberg:2008}. It should be noted that in the monograph by Gorbunov and
Rubakov \cite{Rubakov:2017} another values are taken for the normalized
vacuum density $\Omega _{\Lambda }\approx 0.76$ and for the normalized
(non-relativistic) matter density $\Omega _{M}\approx 0.24$. But in both
cases, the normalization condition $\Omega _{\Lambda }+\Omega _{M}=1$ is
fulfilled which means that $\Omega _{rad}$ should be neglected.
Nevertheless, in \cite{Rubakov:2017} an estimate has been made that the
contributions of ultra-relativistic and non-relativistic matter are comparable
when
\begin{equation}
z_{eq}+1=\frac{a_{0}}{a_{eq}}\sim \frac{\Omega _{M}}{\Omega _{rad}}\sim
10^{4}\text{ \ .}  \label{C23}
\end{equation}%
This means that $\Omega _{M}\sim 10^{-1}$ and $\Omega _{rad}$ is
proportional to $10^{-5}$. Then the temperature of the Universe is $%
T_{eq}=T_{0}(z_{eq}+1)\sim 10^{4}$ $K\sim 1$ $eV$. In fact, first the
temperature $T_{0}$ has been equal to the temperature $T_{0}=2.75$ $K$ of
the relic photons with an energy density
\begin{equation}
\rho _{\gamma ,0}=2.\frac{\pi ^{2}}{30}T_{\gamma
}^{4}=2.55.10^{-10}GeV/cm^{3}\text{, \ }  \label{C23A}
\end{equation}%
according to the Stefan-Boltzmans law. The numerical factor $2$ in Eq. (\ref%
{C23A}) is related to the two polarizations of the photon. Then it will
follow that $\Omega _{\gamma }=5.1.10^{-5}$ \cite{Rubakov:2017}. So the
first case to be investigated for the numerical estimates of $\overline{g}%
_{2}$ and $\overline{g}_{3}$ (expressions Eq. (\ref{C21}) and Eq. (\ref{C22}%
) is when $\Omega _{\gamma }=\Omega _{R}=\Omega _{rad}\sim 10^{-5}$. Then
Eq. (\ref{C21}) and Eq. (\ref{C22}) can be estimated as
\begin{equation}
\overline{g}_{2}\approx 0.67\Omega _{R}+0.33.10^{-4}\approx 0.397.10^{-4}%
\text{ \ \ }.  \label{C24}
\end{equation}%
This Weierstrass invariant $\overline{g}_{2}$ is with a positive sign, but
the second one $\overline{g}_{3}$ is negative for $\Omega _{R}\sim 10^{-5}$
\begin{equation}
\overline{g}_{3}\approx 0.223.10^{-2}.\Omega
_{R}-4.10^{-8}-0.456.10^{-2}\approx -45600.177\times 10^{-7}\text{ \ \ .}
\label{C25}
\end{equation}%
In  absolute value, the second number is $257627$ times greater than the
first one, i.e. $\mid \overline{g}_{3}\mid \gg $ $\overline{g}_{2}$.

\subsubsection{\label{sec:UltralightNeutrinos}The subcase of massive
(ultra-light) neutrinos}

However, after the Universe cools down to temperatures of $1$ $K$ and below,
massless or light neutrinos should exist with masses $m_{\nu }\lesssim 1$ $%
K\sim 10^{-4}$ $eV$. It should be noted that in the Standard model of
elementary particles (see the Weinberg-Salam model in \cite{Ryder:1987}),
even after spontaneous symmetry breaking only electrons, muons ($\mu $) and
gauge bosons acquire masses, but the photons and the neutrinos do not acquire.
Moreover, the contribution of massless neutrinos (and also the
anti-neutrinos) in the normalized ultra-relativistic energy density $\Omega
_{\nu }$ is \ $\Omega _{\nu }=2.\frac{7}{8}.\frac{\pi ^{2}}{30}.\frac{T_{\nu
,0}^{4}}{\rho _{c}}\approx 10^{-5}$ (chapter $7$ of \cite{Rubakov:2017}).
The factor $\frac{7}{8}$ in the Stefan-Boltzman law is due to the fact that
neutrinos are fermions and the neutrino temperature $T_{\nu ,0}$ is related
to the photon temperature $T_{\gamma ,0}$ as $\frac{T_{\gamma ,0}}{T_{\nu ,0}%
}=\left( \frac{11}{4}\right) ^{\frac{1}{3}}\simeq 1.4$ (the proof is based
on the entropy conservation in the electron-photon plasma- see again chapter
$7$ of \cite{Rubakov:2017}). The estimate for $\Omega _{\nu }$ in fact shows
that massless neutrinos do not influence the expansion of the Universe in
the contemporary epoch. However, one can ask the reasonable question: what
happens if one assumes the existence of three types of neutrinos (electron $\nu _{e}$, muon $%
\nu _{\mu }$ and $\tau -$neutrinos), which due to neutrino oscillations can
acquire certain small masses? The most stringent constraint on the sum of
the masses of all the three neutrino types comes from cosmological
considerations - $\sum\limits_{i=1}^{3}m_{\nu _{i}}<1.2-1.0$ $eV$ (Appendix
C in \cite{Rubakov:2017}). Now the total energy density of ultra-relativistic
matter, according to the previously applied Stefan-Boltzman law, \ will be
the sum of the energy densities for photons $\rho _{\gamma }$ and for
neutrinos $\rho _{\nu }$ (Chapters 4 and 5 of \cite{Rubakov:2017})
\begin{equation*}
\rho _{rad}=\rho _{\gamma }+\rho _{\nu }=\left[ 2\frac{\pi ^{2}}{30}%
T_{\gamma }^{4}+3\times 2\times \frac{7}{8}\frac{\pi ^{2}}{30}T_{\nu }^{4}%
\right]
\end{equation*}
\begin{equation}
=\left[ 2+\frac{21}{4}\times \left( \frac{4}{11}\right) ^{\frac{4}{3}}\right]
\times \left( \frac{\pi ^{2}}{30}T_{\gamma }^{4}\right) =1.68\rho _{\gamma }%
\text{ \ ,}  \label{C26}
\end{equation}%
where the second term $3\times 2\times \left( \frac{7}{8}\frac{\pi ^{2}}{30}%
T_{\nu }^{4}\right) $ accounts for the three types of neutrinos (assumed to
exist), the multiplier
$2$ accounts for the two polarizations (one for the neutrino and one for the
corresponding anti-neutrino) and again, the factor $\frac{7}{8}$ is related
to the fact that neutrinos are fermions. Taking into account Eq. (\ref{C26}) and the
estimate $\Omega _{\gamma }=5.1.10^{-5}$, it follows that the contributions
of $\Omega _{\gamma }$ and $\Omega _{rad}$ are comparable \cite{Rubakov:2017}
(because $\Omega _{rad}=\frac{\rho _{rad}}{\rho _{c}}=\frac{1}{1.68}\frac{%
\rho _{\gamma }}{\rho _{c}}=0.5952$, $\Omega _{\gamma }=0.303.10^{-4}$).
Consequently, $\Omega _{rad}$ $\preceq $ $10^{-4}$, unlike the case when $%
\Omega _{\nu }=2.\frac{7}{8}.\frac{\pi ^{2}}{30}.\frac{T_{\nu ,0}^{4}}{\rho
_{c}}\approx 10^{-5}$ and the neutrinos were massless. So if one substitutes
$\Omega _{rad}$ $\approx 10^{-4}$ in Eq. (\ref{C24}) for $\overline{g}_{2}$
and in Eq. (\ref{C25}) for $\overline{g}_{3}$, one can obtain
\begin{equation}
\overline{g}_{2}\approx 0.67.10^{-4}+0.33.10^{-4}\approx 1.10^{-4}\text{ \ \
}  \label{C27}
\end{equation}%
and also
\begin{equation}
\overline{g}_{3}\approx 0.223.10^{-2}.10^{-4}-4.10^{-8}-0.456.10^{-2}\approx
-45598.17\times 10^{-7}\text{ \ \ .}  \label{C28}
\end{equation}%
Consequently, it becomes clear how important it is to have an algorithm for
calculation of elliptic integrals, accounting for very tiny variations of
the constants $\overline{g}_{2}$ and $\overline{g}_{3}$.

\section{\label{sec:ComparEllipticIntegrals}Comparison of elliptic integrals
in different variables and with different Weierstrass invariants}

In this section the method for calculating the Weierstrass invariants $g_{2}$
and $g_{3}$ will be presented, which is based on the combination of two
approaches. In the first approach (see section "Transforming an elliptic integral
in the Legendre form into an elliptic integral in the Weierstrass form"%
) the initial integral $\widetilde{J}_{0}^{(q)}(y)$ was brought to the
Weierstrass form Eq. (\ref{F1}) (multiplied by the conformal factor $\sqrt{a}%
=\sqrt{\frac{9g_{3}}{g_{2}}.\frac{\left( q^{4}-q^{2}+1\right) }{\left(
2q^{4}-5q^{2}+2\right) }}$) by means of the transformation Eq. (\ref{F2}) $x=%
\frac{a}{y^{2}}+b$, as a result of which the integral acquires the form Eq. (%
\ref{F1}). We shall call this representation "equation A". The second
approach is based on the method of "four-dimensional uniformization",
as a result of which the integral is brought to the second representation Eq. (\ref{C3}) - "equation B".
As a result of the two representations, the following two equalities can be written

\begin{equation}
\widetilde{J}_{0}^{(q)}(y)=\int \frac{dy}{\sqrt{(1-y^{2})(1-q^{2}y^{2})}}=-%
\sqrt{a}\int \frac{dx}{\sqrt{4x^{3}-g_{2}x-g_{3}}}=\int \frac{d\sigma }{%
\sqrt{4\sigma ^{3}-\overline{g}_{2}\sigma -\overline{g}_{3}}}\text{ \ \ \ .}
\label{D1}
\end{equation}

Theorem: (proved in the monograph by Hancock \cite{Hancock:1917})The elliptic
integral in the Weierstrass representation Eq. (\ref{F1}) (equation (A) in
Eq. (\ref{D1}) in terms of the "$x$"-variable) can be written in terms of
the variable $\sigma $ as
\begin{equation}
\int \frac{dx}{\sqrt{4x^{3}-g_{2}x-g_{3}}}=-\int \frac{d\sigma }{\sqrt{%
4\sigma ^{3}-g_{2}\sigma -g_{3}}}  \label{D2}
\end{equation}%
after performing the variable transformation
\begin{equation}
x=x_{0}+\frac{(x_{1}-x_{0})(x_{2}-x_{0})}{\sigma -x_{0}}\text{ \ \ \ ,}
\label{D3}
\end{equation}

where $x_{0},x_{1}$ and $x_{2}$ are the roots of the cubic polynomial $%
4x^{3}-g_{2}x-g_{3}=0$.

In the original formulation of the theorem in \cite{Hancock:1917} \ (page $%
29 $) the transformation $t=e_{1}+\frac{(e_{2}-e_{1})(e_{3}-e_{1})}{s-e_{1}}$
has been applied to the integral $\int \frac{dt}{\sqrt{(t-e_{1})}%
(t-e_{2})(t-e_{3})}$. Note however the minus sign in front of the right term
in Eq. (\ref{D2}), which is important and can easily be confirmed by a
direct calculation. Applying again the above theorem with respect to
equation (B) in Eq. (\ref{D1}) and also Eq. (\ref{C3}), it can be written
\begin{equation}
\int \frac{d\sigma }{\sqrt{4\sigma ^{3}-g_{2}\sigma -g_{3}}}=-\frac{1}{\sqrt%
[3]{a}}\int \frac{d\overline{\sigma }}{\sqrt{4\overline{\sigma }^{3}-%
\widehat{g}_{2}\overline{\sigma }-\widehat{g}_{3}}}\text{ \ \ \ \ ,}
\label{D4}
\end{equation}%
where the variable transformation \ $\sigma =\frac{1}{\sqrt[3]{a}}\overline{%
\sigma }$ has been performed and the notations $\widehat{g}_{2}\equiv $ $%
\overline{g}_{2}a^{\frac{2}{3}}$\ \ \ \ , \ $\widehat{g}_{3}\equiv \overline{%
g}_{3}a$ are introduced. From (A) Eq. (\ref{D2}) and (B) Eq. (\ref{D4}) we
obtain
\begin{equation}
\int \frac{d\sigma }{\sqrt{4\sigma ^{3}-g_{2}\sigma -g_{3}}}=\frac{1}{\sqrt[%
3]{a}}\int \frac{d\sigma }{\sqrt{4\sigma ^{3}-\widehat{g}_{2}\sigma -%
\widehat{g}_{3}}}\text{ \ \ \ ,}  \label{D5}
\end{equation}%
meaning that if the roots of the polynomial in the denominator of the
left-hand side are $(\sigma _{0},\sigma _{1},\sigma _{2})$, then the roots
in the right-hand side should be $(\sigma _{0},\sigma _{1},\sigma _{2})=\sqrt[3]{a}%
(\sigma _{0},\sigma _{1},\sigma _{2})$. From the Cardano formulae, we have
for the roots of the cubic polynomial $4\sigma ^{3}-g_{2}\sigma -g_{3}=0$ \ (%
$i=1,2,3$)
\begin{equation}
\sigma _{i}=\sqrt[3]{\frac{g_{3}}{8}+\frac{1}{4}\sqrt{\frac{g_{3}^{2}}{4}-%
\frac{1}{27.4}g_{2}^{3}}}+\sqrt[3]{\frac{g_{3}}{8}-\frac{1}{4}\sqrt{\frac{%
g_{3}^{2}}{4}-\frac{1}{27.4}g_{2}^{3}}}\text{ \ \ .}  \label{D6}
\end{equation}%
These roots should be equal to $\sqrt[3]{a}(\sigma _{0},\sigma _{1},\sigma
_{2})$, where $4\sigma ^{3}-\widehat{g}_{2}\sigma -\widehat{g}_{3}=0$
\begin{equation}
\sqrt[3]{a}\sigma _{i}=\sqrt[3]{-\frac{a\widehat{g}_{3}}{8}+\frac{1}{4}\sqrt{%
\frac{a^{2}\widehat{g}_{3}^{2}}{4}-\frac{1}{27.4}a^{2}\widehat{g}_{2}^{3}}}+%
\sqrt[3]{-\frac{a\widehat{g}_{3}}{8}-\frac{1}{4}\sqrt{\frac{a^{2}\widehat{g}%
_{3}^{2}}{4}-\frac{1}{27.4}a^{2}\widehat{g}_{2}^{3}}}\text{ \ \ .}
\label{D7}
\end{equation}%
From the equality of the two systems of roots, given by Eq. (\ref{D6}) and
Eq. (\ref{D7}) it follows that
\begin{equation}
g_{2}^{3}=a^{2}\widehat{g}_{2}^{3}\text{ \ \ \ , \ \ \ \ \ \ }g_{3}=-a%
\widehat{g}_{3}\text{ \ \ \ \ .}  \label{D8}
\end{equation}%
Taking into account the equality 
\begin{equation}
\widehat{g}_{2}\equiv \overline{g}_{2}a^{\frac{2}{3}}\ \ \ \ ,\ \widehat{g}%
_{3}\equiv \overline{g}_{3}a  \label{D9}
\end{equation}%
and also the definition for $a\equiv \frac{3g_{3}}{g_{2}}K$, we can obtain
the important relations
\begin{equation}
g_{2}=\overline{g}_{2}a^{\frac{4}{3}}=(3K)^{4}\frac{\overline{g}_{3}^{4}}{%
\overline{g}_{2}^{3}}\text{ \ \ , \ \ }g_{3}=-a^{2}\overline{g}%
_{3}=-27^{2}K^{6}\frac{\overline{g}_{3}^{7}}{\overline{g}_{2}^{6}}\text{ \ \
\ ,}  \label{D10}
\end{equation}%
which depend in a complicated way on the modulus parameter $q$.

It should be stressed that the formulaes Eq. (\ref{D10}) can be applied not
only with respect to the integral Eq. (\ref{F1}) with the Weierstrass
invariants $\overline{g}_{2}$ and $\overline{g}_{3}$, determined by the
formulaes Eq. (\ref{C8}), Eq. (\ref{C9}) - Eq. (\ref{C10}) (which are
polynomial functions of the modulus parameter $q$), but also they can be applied with respect to
the elliptic integral Eq. (\ref{B5}) for the apparent luminosity distance
with invariants $\overline{g}_{2}$ and $\overline{g}_{3}$. These invariants are  given by
formulaes Eq. (\ref{C21}) and Eq. (\ref{C22}) \ and they are dependent on the numerical parameters $%
\Omega _{\Lambda }$, $\Omega _{R}$, $\Omega _{K}$ and $\Omega _{M}$.
 Equalities A and B in Eq. (\ref{D1})
will also be valid, but the "conformal factor" $\sqrt{a}$ will be given by
another complicated expression. This will be shown in a subsequent
publication.

\subsection{\label{sec:TwoModuli}Can the two moduli $k$ and $\widetilde{k}$
be defined correctly?}

The two moduli are defined by the corresponding expressions 
\begin{equation}
k^{2}=\frac{1}{2}\pm \frac{1}{2}\sqrt{-3+\frac{12}{\sqrt[3]{4}}\overline{g}%
_{2}^{(q)}}\text{ \ \ ,}  \label{D11}
\end{equation}%
\begin{equation}
\widetilde{k}^{2}=\frac{1}{2}\pm \frac{1}{2}\sqrt{-3+\frac{12}{\sqrt[3]{4}}%
g_{2}^{(q)}\text{ \ \ .}}  \label{D12}
\end{equation}%
The first modulus is derived as a solution of the quartic algebraic equation
\begin{equation}
\frac{1}{\sqrt[3]{4}}=\frac{k^{4}-k^{2}+1}{3\overline{g}_{2}^{(q)}}\text{ \
\ }  \label{D12A1}
\end{equation}%
and the second modulus $\widetilde{k}$ has been defined in Eq. (\ref{F18})
as $\widetilde{k}=-1-\frac{6\alpha }{\sqrt[3]{2}}$ \ , where \ $\alpha =%
\widetilde{b}_{2}$ is a root of \ the quartic (cubic) equation $4\widetilde{b}%
_{2}^{3}-g_{2}^{(q)}\widetilde{b}_{2}-g_{3}^{(q)}=0$. The modulus $%
\widetilde{k}$ satisfies also the equality
\begin{equation}
\frac{1}{\sqrt[3]{4}}=\frac{\widetilde{k}^{4}-\widetilde{k}^{2}+1}{%
3g_{2}^{(q)}}\text{ \ }\ .\text{\ }  \label{D12A2}
\end{equation}%
The correct definition implies also that the two inequalities
\begin{equation}
0<k^{2}<1\text{ \ \ \ , \ \ }0<\widetilde{k}^{2}<1\text{ \ \ \ }  \label{D13}
\end{equation}%
should be fulfilled. For the first case, after some lengthy calculations it
can be proved that the first inequality is fulfilled when
\begin{equation}
\sqrt{-7+\sqrt{48+3\sqrt[3]{4}}}\leq q<\sqrt{-7+\sqrt{48+4\sqrt[3]{4}}}<1%
\text{ \ \ \ ,}  \label{D14}
\end{equation}%
so it is correctly defined but in a very narrow range, because the two
numbers in the left-hand and in the right-hand side of the above inequality
are very close one to another. That is why this case should be discarded.

For the second case of the equation Eq. (\ref{D12}), \ the inequality \ $0<%
\widetilde{k}^{2}<1$ should be determined from the more complicated
algebraic equation with respect to the two moduli $\widetilde{k}$ and $q$
\begin{equation}
\widetilde{k}^{4}-\widetilde{k}^{2}+1=\frac{3^{5}}{\sqrt[3]{4}}\left[ \frac{%
q^{4}-q^{2}+1}{2q^{4}-5q^{2}+2}\right] ^{4}.\frac{\overline{g}_{3}^{4}(q)}{%
\overline{g}_{2}^{3}(q)}\text{ \ \ .}  \label{D15}
\end{equation}

The interchange $q\Leftrightarrow \widetilde{k}$ gives the other complicated
algebraic equation
\begin{equation}
q^{4}-q^{2}+1=\frac{3^{5}}{\sqrt[3]{4}}\left[ \frac{\widetilde{k}^{4}-%
\widetilde{k}^{2}+1}{2\widetilde{k}^{4}-5\widetilde{k}^{2}+2}\right] ^{4}.%
\frac{\overline{g}_{3}^{4}(\widetilde{k})}{\overline{g}_{2}^{3}(\widetilde{k}%
)}\text{ \ \ .}  \label{D16}
\end{equation}%
The simultaneous fulfillment of these two algebraic relations Eq. (\ref{D15}%
) and Eq. (\ref{D16}) does not impose serious restrictions on $q$ and $%
\widetilde{k}$.

\section{Conclusion}

\bigskip In this paper two major tasks have been carried out. 1. Several
types of elliptic integrals have been pointed out, which have various
physical important applications. Regretfully, most papers, dedicated to elliptic integrals in various 
mathematical problems, apply predominantly numerical approaches. The present paper demonstrates that there are also 
analytical approaches for solution of elliptic integrals, which are not standard and need further investigation. 
Some of the basic monographs on elliptic
integrals such as \cite{Legendre:1825}, \cite{WhittakerWatson:1927},
\cite{Whittaker:1947}, \cite{Timofeev:1948}, \cite{Smirnov3}, \cite{Hancock:1917}
are very old, but contain the basic approaches. Contemporary monographs such
as \cite{PrasSol:1997} contain very good pedagogical introductions to the
topic of elliptic curves and integrals, but some not so standard topics such
as the s.c. "method for four-dimensional uniformization" remain outside the
scope of such monographs. Nevertheless, in older monographs such as \cite%
{Whittaker:1947} this method is exposed in details. 2. The method of
"four-dimensional uniformization" turned out to be very useful for
transforming an initial "four-dimensional" elliptic integral into an
integral in the Weierstrass representation and also for comparing the two elliptic integrals in
Eq. (\ref{D1}). This by itself constitutes a separate mathematical problem,
namely: the first integral in the Legendre form in Eq. (\ref{D1}) is equal to another elliptic
integral in the Weierstrass form, "conformally" multiplied by the factor $%
\sqrt{a}$, where $a$ is determined in Eq. (\ref{F3}) as $a=\frac{9g_{3}}{%
g_{2}}K(q)$ and $K(q)$ $\equiv \frac{q^{4}-q^{2}+1}{2q^{4}-5q^{2}+2}$ in Eq.
(\ref{F5}). Due to the presence of higher degrees of the modulus parameter $%
q $ in the expression for $K(q)$, it might seem that applying the
transformation Eq. (\ref{F2}) $x=\frac{a}{y^{2}}+b$ is not very appropriate
because the eccentricity of the GPS\ orbit $e=0.01323881349526$ is a very
small number (according to the PhD thesis \cite{Gulklett:C4}). However,
modern technologies for space research are based on atomic clocks with cold
atoms, mounted on space platforms on highly elliptical orbits, so that the
atomic clocks will be very sensitive to the varying potential of the
gravitational field in the near-Earth space \cite{Alonso:2022}. That is why,
the effects of the ellipticity of the orbit should not be neglected, even
if it may be a very small number. It should be clarified that the
eccentricity $e$ of the GPS orbit plays the role of the modulus parameter $q$ in the
expression for the propagation time of the signal (see \ \cite{DimitrovAIP:2022}).

One more important peculiarity of the approach in this paper should be
mentioned. In the monographs \cite{Fichten2} and \cite{Smirnov3} the
recurrent system of equations for the $3D$ case (in terms of the $x-$%
variable) has been studied, while in the monograph \cite{PrasSol:1997}-for
the $4D$ case (in terms of the $y-$variable). The two variables of course
are related one with another by means of the transformation Eq. (\ref{F2}).
In this paper for the first time the relation between the two recurrent
systems has been studied, which gave us also the opportunity to derive the
relation Eq. (\ref{D1}) - the integrals "A" and "B". The second integral "B" is obtained as a result of
the method of "four-dimensional" uniformization, at first exposed in the
monograph \cite{WhittakerWatson:1927} and subsequently - applied in this paper.
It is remarkable, however,  that from the integral "B" in Eq. (\ref{D1}) one can obtain again the original
integral in the Legendre representation after applying the transformations $%
x=\frac{\widetilde{a}_{2}}{y^{2}}+\widetilde{b}_{2}$ and $y\Rightarrow \frac{%
1}{\widetilde{k}y}$, which give the opportunity for introducing the second
modulus \ $\widetilde{k}$ in the resulting transformation Eq. (\ref{F16}) $%
\widetilde{x}\equiv \widetilde{a}_{2}\widetilde{k}^{2}y^{2}+\widetilde{b}%
_{2} $. Due to the "mirror" symmetry of
 Eq. (\ref{F19A}) with respect to the moduli $q$ and $\widetilde{k}$ and the
possibility to perform the interchange $q\Leftrightarrow \widetilde{k}$ in
Eq. (\ref{F23}), it became possible to obtain the final Eq. (\ref{F24}) for
the zero-order elliptic integral $J_{0}^{(q)}(y)$.

\begin{acknowledgments}
The author is grateful to the organizers of the Fourteenth Hybrid Conference of the
Euro-American Consortium for Promoting  the Application of Mathematics in Technical and
Natural Sciences   (AMITaNS’22) 22-27 June 2022, Albena resourt, Bulgaria and especially to
Prof. Michail Todorov for the opportunity to participate in this conference. The author is grateful also to
the Institute for Advanced Physical Studies (IAPS) for some financial support, and also to the anonymous referee 
for his/her constructive critical remarks.

\end{acknowledgments}

% References

\end{document}